\documentclass[10pt,letterpaper]{article}
\usepackage{opex3}
\usepackage{tikz}
\usepackage{amsmath,amssymb}

\newcommand{\R}{\mathbb{R}}
\newcommand{\norm}[1]{|| #1 ||}

\begin{document}

\title{Reconstruction of hidden 3D shapes using diffuse reflections}
\author{Otkrist Gupta$^1$, Andreas Velten$^2$, Thomas Willwacher$^3$, Ashok Veeraraghavan$^4$, and Ramesh Raskar$^1$\\$^1$MIT Media Lab, $^2$University of Wisconsin Madison, $^3$Harvard University, $^4$Rice University}


%
\newenvironment{tight_itemize}{\begin{itemize} \itemsep 
-2pt}{\end{itemize}}
\newenvironment{tight_enumerate}{\begin{enumerate} \itemsep 
-2pt}{\end{enumerate}}







\begin{abstract}
We analyze multi-bounce propagation of light in an unknown hidden volume and demonstrate that the reflected light contains sufficient information to recover the 3D structure of the hidden scene. We formulate the forward and inverse theory of secondary and tertiary scattering reflection using ideas from energy front propagation and tomography. We show that using careful choice of approximations, such as Fresnel  approximation, greatly simplifies this problem and the inversion can be achieved via a backpropagation process. We provide a theoretical analysis of the invertibility, uniqueness and choices of space-time-angle dimensions using synthetic examples. We show that a 2D streak camera can be used to discover and reconstruct hidden geometry. Using a 1D high speed time of ﬂight camera, we show that our method can be used recover 3D shapes of objects "around the corner".\end{abstract}


\bibliographystyle{osajnl}


%
%
%
%
%
%





%



\section{Introduction}
Recent work in Light Transport (LT) in optics, computer graphics and computer vision has shown ability to recover surprising geometric and photometric information about the scene from diffuse scattering. 
The pioneering work in Dual Photography \cite{Sen05} shows one can exploit the second bounce to recover images of hidden objects. The theory of inverse light transport \cite{Seitz05} can be used to eliminate inter-reflections from real scenes.
The frequency domain properties of direct and global components of scattered light can be exploited to recover images of objects behind a shower curtain \cite{Nayar06}.
Three bounce analysis of a time-of-flight camera can recover hidden 1-0-1 barcodes \cite{iccv09femto} and single object position \cite{pandharkar11} while form factors can be recovered from just two bounces \cite{liu2010shape}. The recent paper \cite{velten11} presents an approach to recover 3D shape from tertiary diffuse scattering.

Similar to these and other inverse light transport approaches \cite{Seitz05}, we illuminate one spot at a time with our pulsed laser projector and record the reflected light after its interaction with the scene.
We record an extra time dimension of light transport with an imaging system that uses a short duration pulse and a time-of-flight camera. A schematic of our setup can be found in Figure \ref{fig:notations}. 
We show that this extra temporal dimension of the observations makes the 3D structure of a hidden scene observable.
The lack of correspondence between scene points and their contributions to the captured streak images after multiple bounces is the main challenge and we show computational methods to invert the process.


\paragraph{Contributions}
We explore the relationship between the hidden 3D structure of objects and the associated high dimensional light transport (space and time). We show that using a multi-bounce energy front propagation based analysis one can recover hidden 3D geometry. We analyze the problem of recovering a 3D shape from its tertiary diffuse reflections. If there was only a single hidden point, the reflected energy front directly encodes the position of that point in 3D. We rigorously formulate the problem, elicit the relationships between geometry and acquired light transport and also develop a practical and robust framework for inversion. We show that it can be cast as a very peculiar type of tomographic reconstruction problem. We call the associated imaging process \emph{elliptic tomography}. The inverse problem, i.e., the recovery of the unknown scene from the measurements, is challenging. We provide analysis of a fast algorithm, which is essentially the analogue of the filtered backprojection algorithm in traditional tomography. We perform several synthetic and physical experiments to validate the concepts.


\paragraph{Limitations and Scope}
Our multi-scattering based method is inherently limited by the signal to noise ratio due to serious loss in scattering. 
We require fine time-resolution and good sensitivity. Scenes with sufficient complexity in geometry (volumetric scattering, large distances) or reflectance (dark surfaces, mirrors) can pose problems. We limited our scope to scenes with approximately Lambertian reflectance (no directionally varying BRDF) 
and with few or no partial occlusions for simplicity. In order to achieve  good spatial resolution, we require the visible area to be sufficiently large and the object to be sufficiently close to this visible area, so that we have a large 'baseline' to indirectly view hidden objects.
%
%
\section{Modeling Propagation of a Light Pulse for Multiple Bounces}

\begin{figure}[ht]
\centering
\includegraphics[height=2.0in]{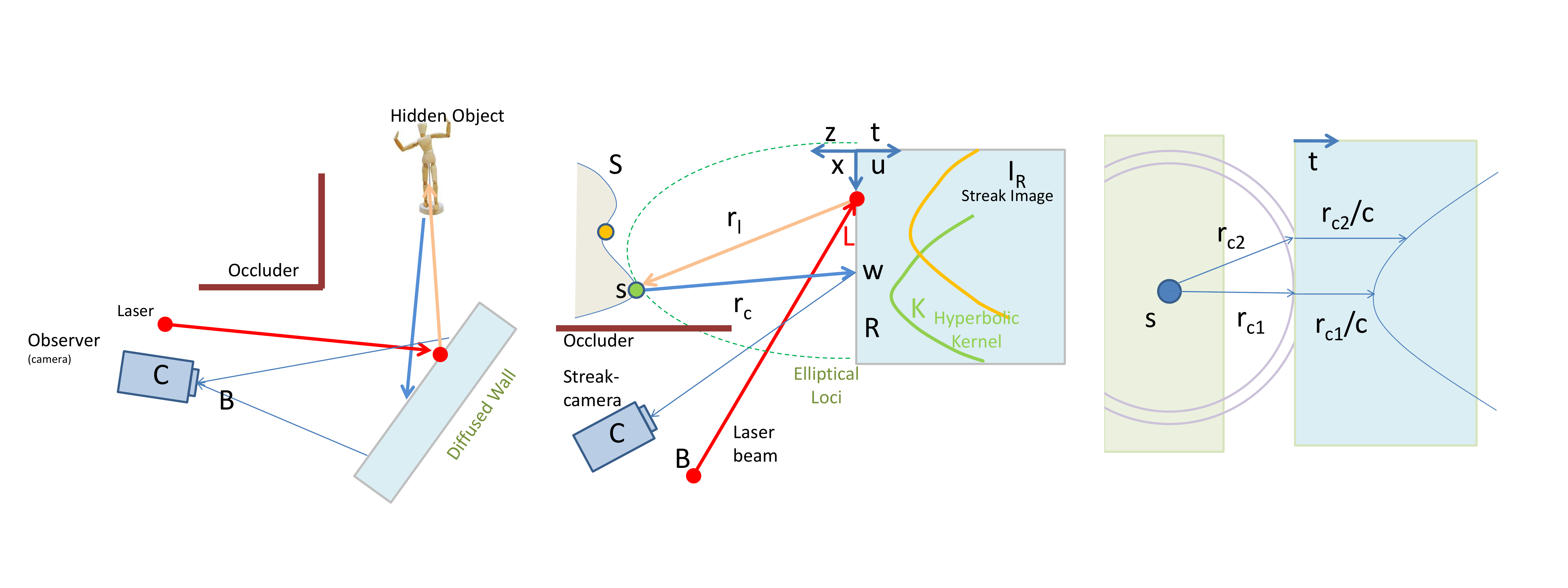}
\caption{\label{fig:notations} Forward Model. (Left) The laser illuminates the surface S and each point $s \in S$ generates a energy front. The spherical energy front contributes to a hyperbola in the space-time streak photo, $I_R$. \label{fig:hyperbola} (Right) Spherical energy fronts propagating from a point create a hyperbolic space-time curve in streak photo.}
\end{figure}

{To develop a method for reconstruction of images from time-of-flight information we need to obtain an adequate, physically realistic model of light propagation in the scene. The model can then be used to understand the forward process of rendering time-of-flight streak camera images from a known geometry and finally the inverse process of reconstructing the hidden geometry from the images. The light transport model is sketched in Figure~\ref{fig:notations}. Light emitted by a laser (B) in a collimated beam  strikes a diffuse surface, the {\it diffuser wall} at a single point L. The laser emits pulses that are much shorter than 1~ps in time, or 0.3~mm in space. From point L light is diffusely scattered in all directions. A small fraction of the light travels a distance $r_l$ and strikes point s. At s the light is diffusely reflected once again and some of it travels to point w covering the distance $r_c$. From w some light travels back to the camera C. The camera observes points on the wall with high time resolution, such that light having traveled different total distances through the scene is detected at different times.}

{By scanning the position of L, and exploiting the spatial resolution of the camera we can probe different sets of light paths. 5D light transport captured in this way contains information about the hidden object and allows us to uniquely identify the hidden geometry. Consider for example just one reflecting spot s in the hidden scene as indicated in Figure~\ref{fig:hyperbola}. The space time image of this spot for a given laser position L is a hyperbola. The curvature and position of the hyperbola indicate the position of s. In our experimental setup we make use of a streak camera with a 1 dimensional field of view. We thus only capture 4D light transport. This does, however, still allow us to reconstruct the hidden geometry. The reconstruction quality along the axis perpendicular to the camera field of view is, however, compromised.}

%

\subsection{Space-Time Warping for Bounce Reduction}
\label{subsection:spacetimetransform}
\begin{figure}[ht]
\centering
\includegraphics[width=5.0in]{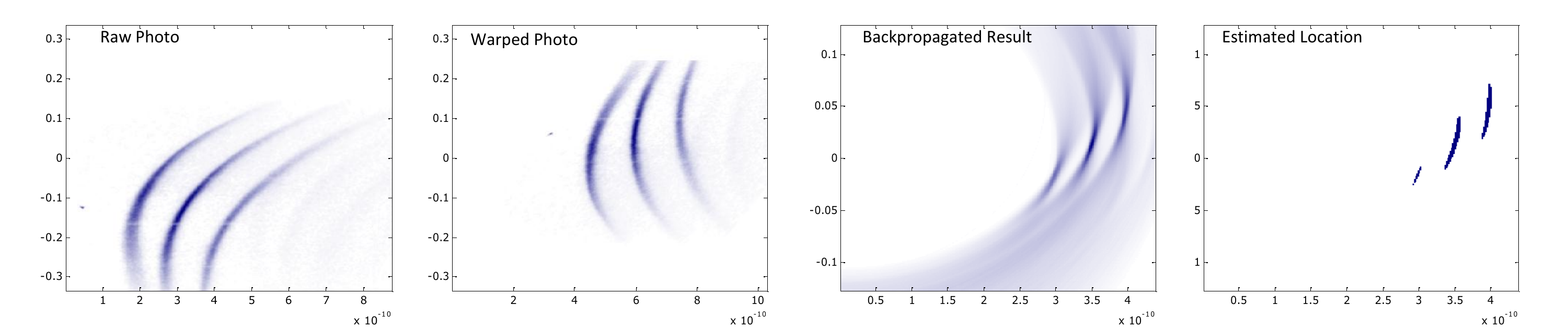}\\
\caption{A space time transform on a raw streak photo allows us to convert a 4 segment problem into a sequence of 2 segment problems. The toy scene is a small 1cm$\times$1cm patch creating a prominent (blurred) hyperbola in thewarped photo. Backprojection creates a low frequency residual but simple thresholding recovers the patch geometry.}
\end{figure}

{The light path from laser to camera can be divided into four straight line segments with three bounces in between. The first segment is the collimated laser beam travelling from the laser source to the wall. In the second segment the laser spot on the wall behaves as a point light source and light travels from there into the hidden scene. The third segment(s) involve scattering from the hidden object. For the fourth segment light travels from the wall to the camera \textit{which is focused on the wall}. The data received by the camera $I_c(p,t)$ has two degrees of freedom - space and time. Since the first and fourth segments are focused, we can apply transforms to the streak images to eliminate their effects. The effect of the first segment can be removed by shifting the streak images in time for each laser location on the wall. To remove the effect of the fourth segment we use the distances between the wall pixels and the camera sensors $(\norm{C-w})$. We assume that we know the geometry of $R$ to calculate the camera-to-wall homography and the correspondence between the camera and wall pixels. The following mathematical formulation is a concise representation of this concept. Here $H$ is the projective transformation (homography) mapping coordinates on $R$ to camera coordinates. The time shift $\norm{C-w}$ by the distance from camera to screen varies hyperbolically with the pixel coordinate $w$. Note that we don't require to adjust for a $\cos(\theta)$ factor or $1/r^2$ fall off because the camera sensor integrates for more wall pixels if they are farther away.}
\begin{equation}
\label{equ:preprocessing}
 I_R(w,t) = I_C(H(w), t  - \norm{L-B} - \norm{C-w}).
\end{equation}

\subsection{Scattering of a pulse}
\label{sec:impulsescattering}

\subsubsection{Generating Streak Photos}
{Let us analyze the scattering of light in the second and third segments. For simplicity we model the hidden object as a collection of unfocused emitters sending impulses $I_s(s,\tau)$ at times $\tau$. We model the receiver as an array of unfocused receivers which capture photons at picosecond resolution. We can achieve this configuration experimentally by using a picoseond resolution streak camera focused on the diffuser wall pixels. 
The following mathematical equation represents the data recorded by the streak camera after making the mentioned assumptions. 
By measuring time in distance units we can set the speed of light to $c=1$ for simplicity. We also ignore the local changes in normals for sender surface and receiver surface.}
\begin{equation}
\label{equ:IR}
I_R(w, t) = \int_S   \int_\tau    \frac{1}{\pi r_c^2} \delta( r_c - t+\tau ) I_S(s, \tau ) d\tau d^2 s
\end{equation}
where 
$w\in R$, $s\in S$, $t,\tau\in \R$ and $r_c = \norm{ w-s } $. 
and
$I_R(w, t)$ is the intensity observed at $w\in R$ at time $t$.
{After removing the time shifts as described in the previous section and applying transforms from the calculated homography we can further simplify the equation and remove the $\delta$ required to adjust for receiver camera distances. The following equation provides a mathematical summary of this analysis. Note that over all these equations we assume that the receiver and sender are perfectly Lambertian and ignore the local variation in normal vectors. }
Equation \eqref{equ:IR} hence becomes
\begin{equation}
\label{equ:IR2}
I_R(w, t) =
\int_{S} I
 \frac{1}{\pi r_c^2}
 \frac{1}{\pi r_l^2}
\delta(t - r_c-r_l)  d^2 s
\end{equation}

\subsubsection{Hyperbolic Contribution}
{Let us analyze the relationship between the time when a sender emits a pulse and the time and location of a receiver detecting the light. For a fixed sender the response function is a hyperboloid in space and time given by the following mathematical equation. The parameters of the hyperboloid depend on location of sender, a lateral displacement leads to shifts, while a displacement in depth corresponds to flattening. Change in sender time equates to a constant time shift for responses to any of the receivers:}
\begin{equation}
\label{equ:hyper}
t- r_l =  r_c  = \sqrt{ (x-u)^2 + (y-v)^2 + z(x,y)^2  }
\end{equation}
where $u$, $v$ are the two coordinates of $w$ in the receiver plane.
{Careful observation shows that this equation describes an ellipsoid in sender location if we fix the laser and receiver location. The ellipsoid's parameters depend on the time at which a receiver receives an impulse. The laser spot and the receiver (on the wall) constitute the two foci of this ellipsoid. The eccentricity depends on the time when the impulse is received.}

\section{Forward model: Elliptical Tomographic Projection}
In this section we rephrase the above approximation to the forward light transport using notions from tomography. In an idealized case, the inverse problem of recovering the hidden shape can be solved explicitly. We use this explicit solution to inspire our algorithm for real world scenarios.

\subsection{Elliptical tomography problem description}
Our problem has similarities to tomographic projection. Let us rewrite equation \eqref{equ:IR2} in the following form
\begin{align*}
I_R(w, t,L) &=
\int_{S} I
 \frac{1}{\pi r_c^2}
 \frac{1}{\pi r_l^2}
\delta(t - r_c-r_l)  d^2 s
= 
\int_{\R^3} 
 \frac{1}{\pi r_c^2}
 \frac{1}{\pi r_l^2}
\delta(t - r_c-r_l)  I\delta_S(x) d^3 x
\\ &=
\int_{\R^3} 
 \frac{1}{\pi r_c^2}
 \frac{1}{\pi r_l^2}
\delta(t - r_c-r_l)  W(x) d^3 x
\end{align*}

where the unknown world volume $W(x)=I\delta_S(x)$ is a delta function with support on the surface $S$ to be reconstructed.
Apart from the $1/r^2$ factors, which typically vary slowly over the region of interest, we hence see that individual streak image pixels measure elliptical projections of the world volume $W(x)$. Due to its similarity with traditional tomography, we call this problem \emph{elliptical tomography}. Note however that there are also key differences to traditional tomography: (i) The recorded projections live in a higher dimensional (5D) space than the world (3D). (ii) The projections are along (2D) ellipsoids instead of (1D) lines. This makes the analysis much more complicated.

\subsection{Challenges and missing cones}
\label{sec:missingcones}
\begin{figure}
\centering
\includegraphics[width=5.5in]{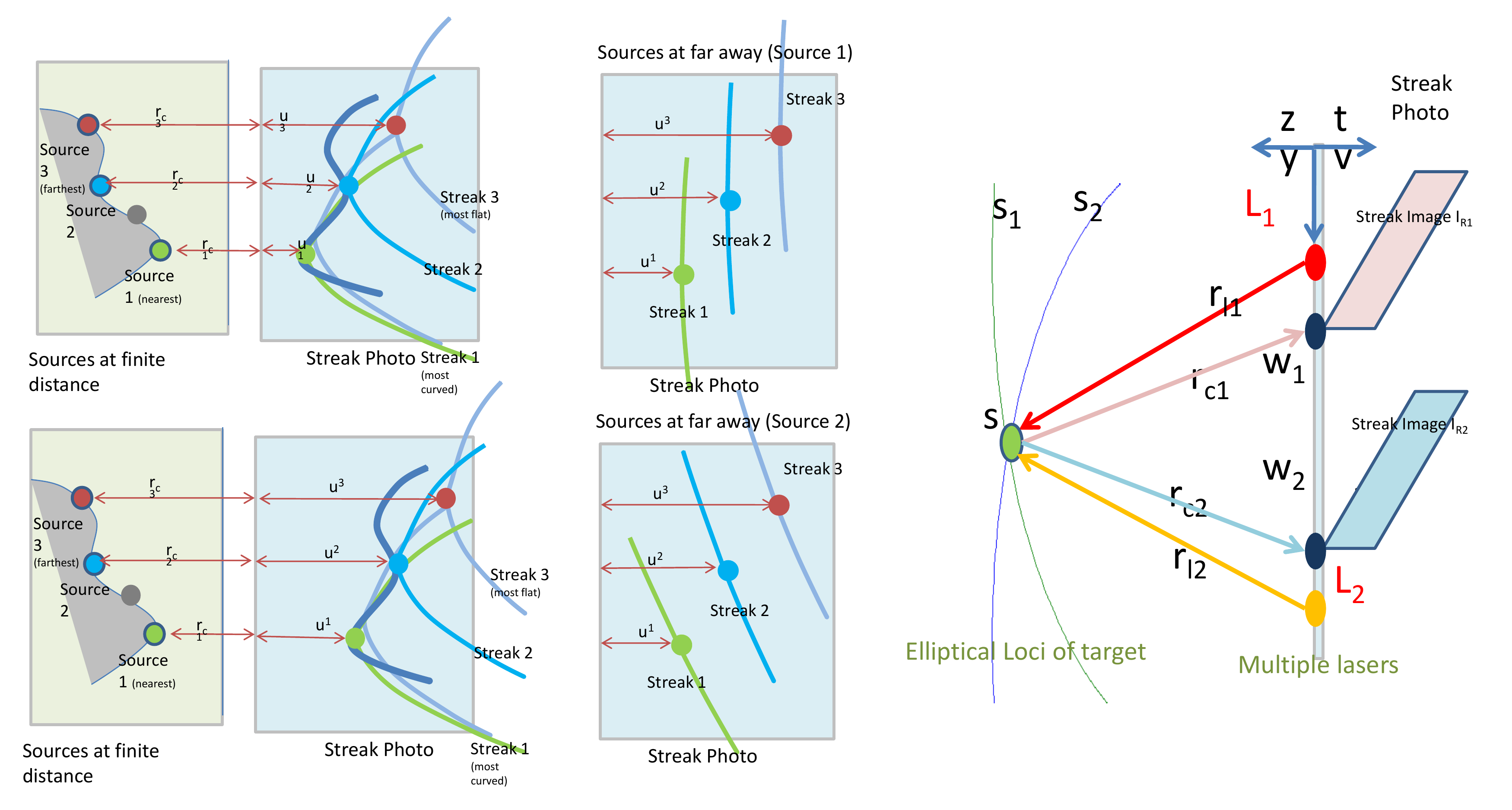}
\caption{\label{fig:tomography} The top left figure shows streak images being generated by near field sources. On bottom left we see effect when this sources travel farther away, The rightmost figure depicts how we can analytically predict single sources using multiple sensor laser combinations. Notice how the accuracy is effected if lasers shift.}
\end{figure}

It is instructive to consider the above tomography problem in the limit when the object is small compared to the distance to the diffuser wall. In this case the elliptic tomography problem reduces to a planar tomography problem, see Figure \ref{fig:tomography}. Each pair of a camera point and a laser position on the diffuser wall (approximately) measures intersections of the target object with planes whose normals agree with the normals to the ellipsoids. By the standard Fourier slice theorem, each line of each streak image will hence measure one line of the Fourier transform of the object, in the direction of the normal. 
Unfortunately, in our situation these normals cover a limited region of the unit sphere. Hence without additional priors it is not possible to reconstruct the Fourier transform of the target object in the missing directions. This is the missing cones problem well known from traditional tomography. Experimentally we get a very good resolution in the depth (orthogonal to the wall) direction, while transverse (parallel to the wall) high frequency features tend to get lost.

\section{Inverse Algorithm: Filtered Back Projection}

In this section we give a detailed description of our reconstruction algorithm.

\subsection{Overview of the algorithm}
The imaging and reconstruction process consists of 3 phases:
\begin{itemize}
\item {\bf Phase 1: Data Acquisition.} We direct the laser to 60 different positions on the diffuser wall and capture the corresponding streak images. For each of the 60 positions XX images are taken and overlayed to reduce noise.
\item {\bf Phase 2: Data Preprocessing.} The streak images are loaded, intensity corrected and shifted to adjust for spatiotemporal jitter.
\item {\bf Phase 3: 3D Reconstruction.} The clean streak images are used to reconstruct the unknown shape using our backprojection-type algorithm.
\end{itemize}

The first of the three phases has been described above.
Let us focus on Phases 2 and 3.

\subsection{Phase 2: Data Preprocessing}

\begin{enumerate}
\item {\bf Timing correction.} To correct for drift in camera timing synchronization (jitter) both in space and time we direct part of the laser directly to the diffuser wall. This produces a sharp ``calibration spot'' in each streak image. The calibration spot is detected in each image, and the image is subsequently shifted in order to align the reference spot at the same pixel location in all streak images.
The severity of the jitter is monitored in order to detect outliers or broken datasets.

\item {\bf Intensity correction.} To remove a common bias in the streak images we subtract a reference (background) image taken without the target object being present in the setup.

\item {\bf Gain correction.} We correct for non-uniform gain of the streak camera's CCD sensor by dividing by a white light image taken beforehand.

\end{enumerate}

\subsection{Phase 3: 3D Reconstruction}

\begin{enumerate}
\item {\bf Voxel Grid Setup.} We estimate an oriented bounding box for the working volume to set up a voxel grid (see below).

\item {\bf Downsampling (optional).} In order to improve speed the data may be downsampled by discarding a fraction of the cameras pixels for each streak image and/or entire streak images. Experiments showed that every second camera pixel may be discarded without losing much reconstruction accuracy. When discarding entire streak images, is is important that the laser positions on the diffuser wall corresponding to the remaining images still cover a large area. 


\item {\bf Backprojection.} For each voxel in the working volume and for each streak image, we compute the streak image pixels that the voxel under consideration might have contributed. Concretely, the voxel at location $v$ can contributed to a pixel corresponding to a point $w$ on the wall at time $t$ if 
\[
ct = |v-L| + |v-w| +|w-C|.
\]
Here $C$ is the camera's center of projection and $L$ is the laser position as above. Let us call the streak image pixels satisfying this condition the \emph{contributing pixels}. We compute a function on voxel space, the \emph{heatmap} $H$. For the voxel $v$ under consideration we assign the value
\[
H(v) = \sum_p (|v-w||v-L|)^\alpha I_p.
\]
Here the sum is over all contributing pixels $p$, and $I_p$ is the intensity measured at that pixel. The prefactor corrects for the distance attenuation, with $\alpha$ being some constant. We use $\alpha=1$.

\item {\bf Filtering.} The heatmap $H$ now is a function on our 3 dimensional voxel grid. We assume that the axis of that grid are ordered such that the third axis faces away from the diffuser wall. We compute the \emph{filtered heatmap} $H_f$ as the second derivative of the heatmap along that third axis of the voxel grid.
\[
H_f = - (\partial_3)^2 H.
\]
The filtered heatmap measures the confidence we have that at a specific voxel location there is a surface patch of the hidden object.

\item {\bf Thresholding.}
We compute a sliding window maximum $M_{loc}$ of the filtered heatmap $H_f$. 
Typically we use a 20x20x20 voxel window.
Our estimate of the 3D shape to be reconstructed consists of those voxels that satisfy the condition
\[
H_f > \lambda_{loc} M_{loc} + \lambda_{glob} M_{glob}
\]
where $M_{glob}=\max(H_f)$ is the global maximum of the filtered heatmap and $\lambda_{loc}$, $\lambda_{glob}$ are constants. Typically we use $\lambda_{loc}=0.45$, $\lambda_{glob}=0.15$

\item {\bf Compressive Reconstruction}

We use techniques like SPGL1, and CoSAMP~\cite{needell2009cosamp} as an alternative to back projection and filtering. We rely on the fact that the 3D voxel grid is sparsely filled, containing surfaces which can occupy only one voxel in depth. Since SPGL1 uses only matrix vector multiplications.

\item {\bf Rendering.} The result of thresholding step is a 3D point cloud.
 We use the Chimera rendering software to visualize this point cloud.

\end{enumerate}

In order to set up the voxel grid in the first step we run a low resolution version of the above algorithm with a conservative initial estimate of the region of interest. In particular, we greatly down sample the input data. By this we obtain a low resolution point cloud which is a coarse approximation to the object to be reconstructed. We compute the center of mass and principal axis of this point cloud. The voxel grid is then fit so as to align with these axis.

\subsection{A remark about the filtering step}
To motivate our choice of filter by taking the second derivative, let us consider again the planar (``far field'') approximation to the elliptical tomography problem as discussed in section \ref{sec:missingcones}. In this setting, at least for full and uniform coverage of the sphere with plane directions, the theoretically correct filtering in the filtered backprojection algorithm is a $|k|^2$ filter in Fourier space. In image space, this amounts to taking a second derivative along each scanline. This motivates our choice of filter above. We tested both filtering by the second derivative in image and world space and found that taking the derivative approximately along the target surfaces normal yielded the best results.

\begin{figure}
\centering
\includegraphics[width=5.5in]{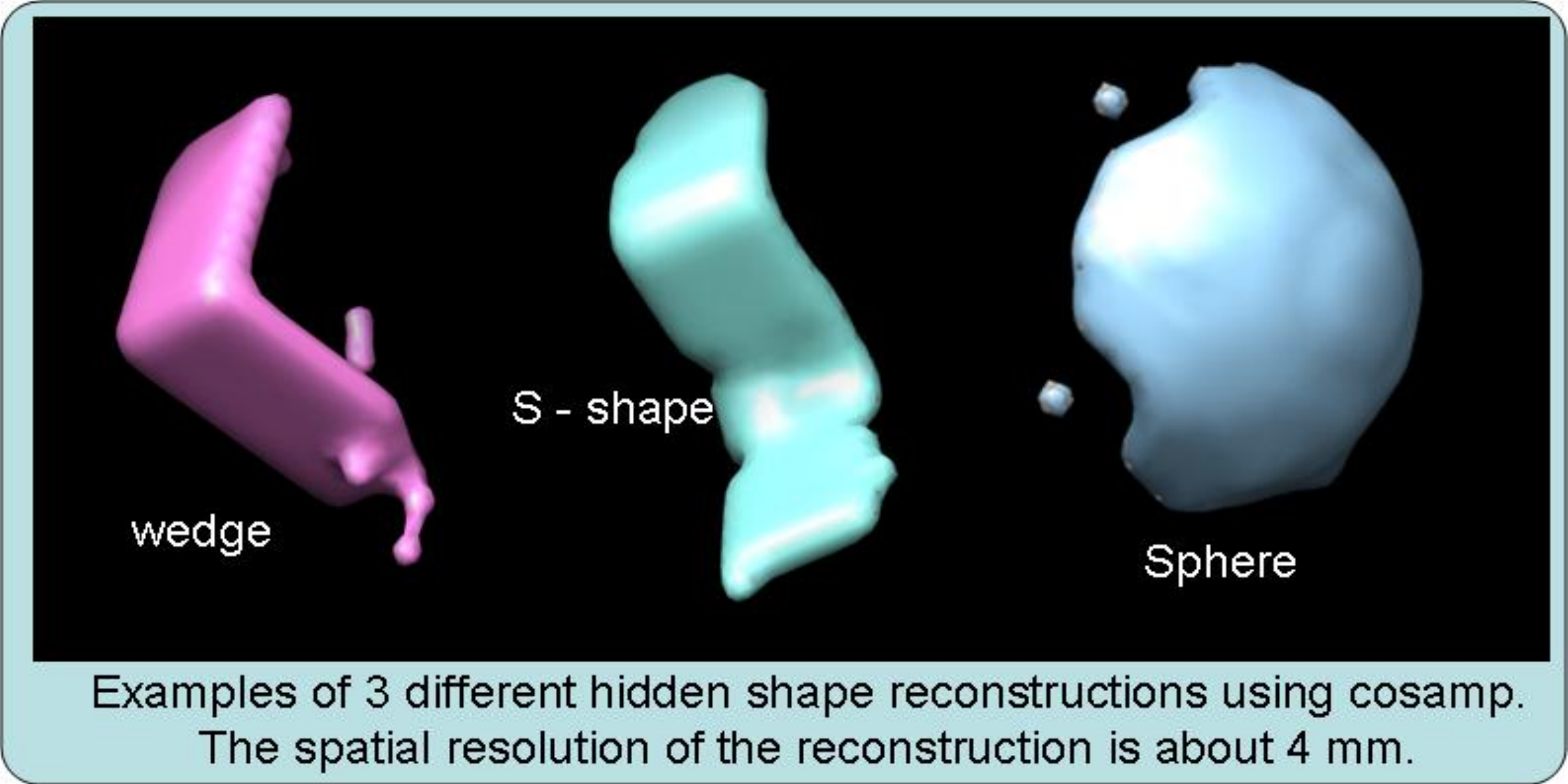}
\caption{\label{fig:cosamp}Simulated reconstruction using CoSAMP. While CoSAMP promises to perform far superior on a perfectly calibrated system, it is outperformed by backprojection on the current data due to calibration inaccuracies.}
\end{figure}

\subsection{Application of CoSAMP}
We attempted compressive reconstructions with CoSAMP on real as well as simulated data. We found that CoSAMP performs better than backprojection on simulated data. The situation is reversed with data from the actual system. We attribute the poor performance of CoSAMP and other linear equation based methods with real data to bias in the data due to imperfect calibration in the system. A result of a simulated CoSAMP reconstruction is shown in Figure~\ref{fig:cosamp}%
%
%
%

\section{Experiments}
\label{sec:experiments}
Our streak camera is a Hamamatsu C5680, with an internal time resolution of 2~picoseconds. We use a mode-locked Ti:Sapphire laser to generate pulses at 795~nm wavelength with about 50~femtosecond pulse duration at a repetition rate of 75~MHz. The laser's average output power is about 500~mW. The streak camera has a one dimensional field of view, imaging a line in the scene. It provides a two dimensional image in which one dimension corresponds to space and the other to time~\cite{hamamatsu_tutorial}. 

We use a Faro Gauge measurement arm to calibrate the laser and camera. We treat the camera and laser as a rigid pair with known intrinsic and extrinsic parameter \cite{forsyth}. The visible parts of the geometry could be measured with the laser directly using time of flight with micrometer precision. Methods like LiDAR and OCT can achieve this and are well understood. In the interest of focusing on the novel components of our system we instead measure the visible parts of our scene with the Faro Gauge. We also collect ground truth data to validate our reconstructions.

\subsection{Results}
\begin{figure}
\centering
\includegraphics[height=1in]{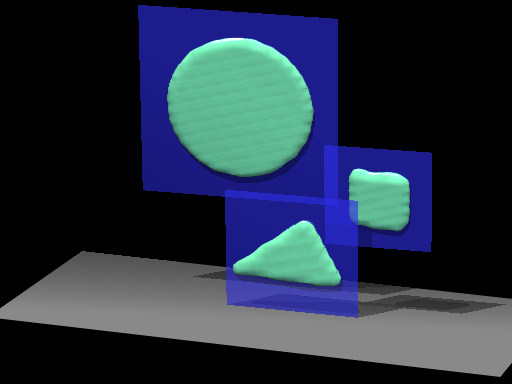}
\includegraphics[height=1in]{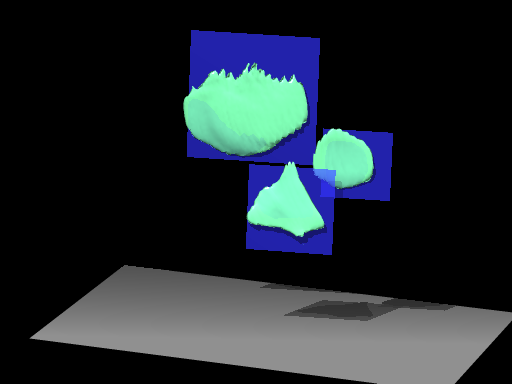}
\includegraphics[height=1in]{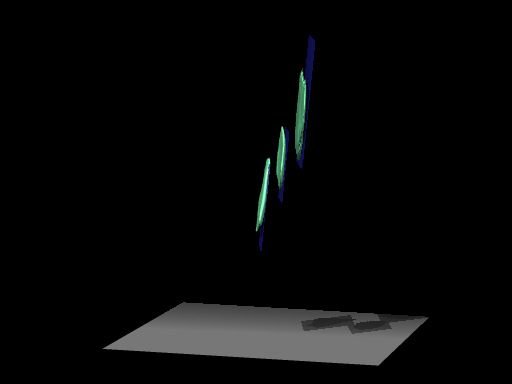}

\caption{\label{fig:recon_shapes} Reconstruction of a scene consisting of a big disk, a triangle and a square at different depth. (Left) Ground truth.
(Middle) Reconstruction, front view. (Right) Reconstruction, side view.
Note that the disk is only partially reconstructed, and the square is rounded of, while the triangle is recovered very well. This illustrates the diminishing resolution in directions parallel to the receiver plane towards the borders of the field of view. The blue planes indicate the ground truth. The gray ground planes and shadows have been added to help visualization. }
\end{figure}

\begin{figure}
\centering

\includegraphics[height=1in]{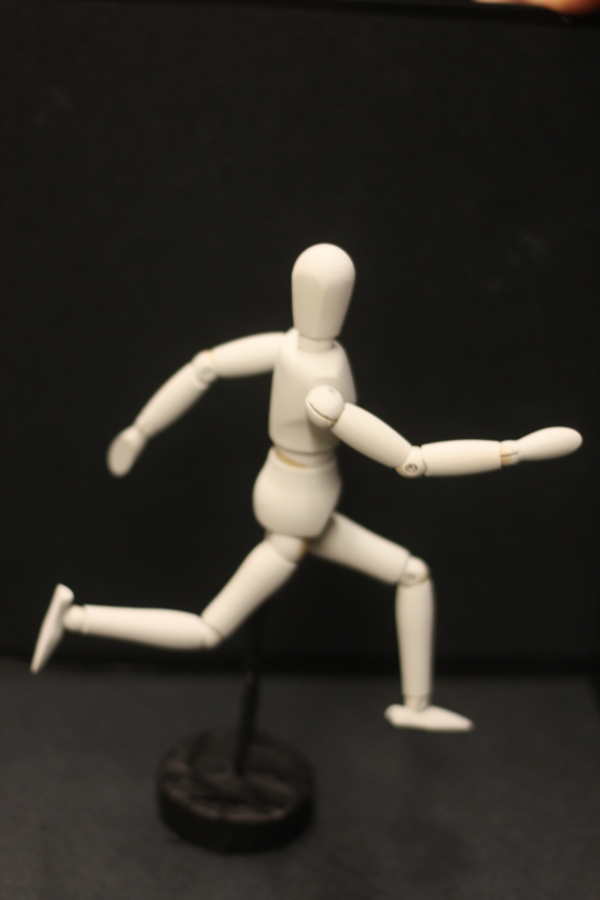}
\includegraphics[height=1in]{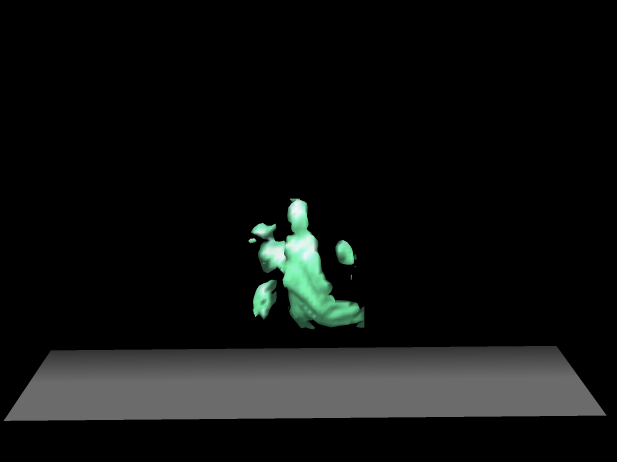}
\includegraphics[height=1in]{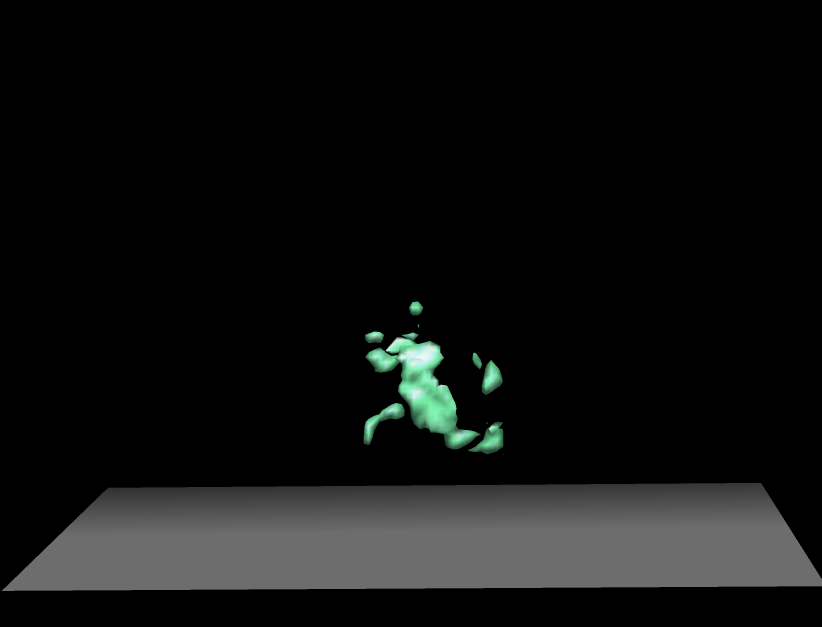}

\caption{\label{fig:recon_gandalf} Reconstruction of a wooden man, painted white. Center - reconstruction using simple back projection based methods. Right - reconstruction using sparse reconstruction methods.  }
\end{figure}

\begin{figure}
\centering

\includegraphics[height=1in]{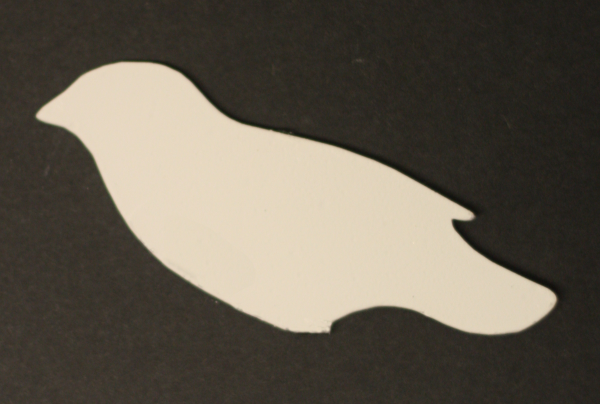}
\includegraphics[height=1in]{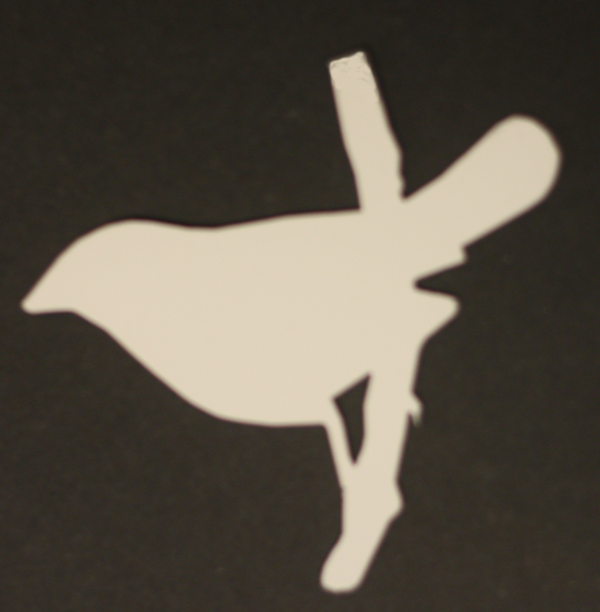}
\includegraphics[height=1.3in]{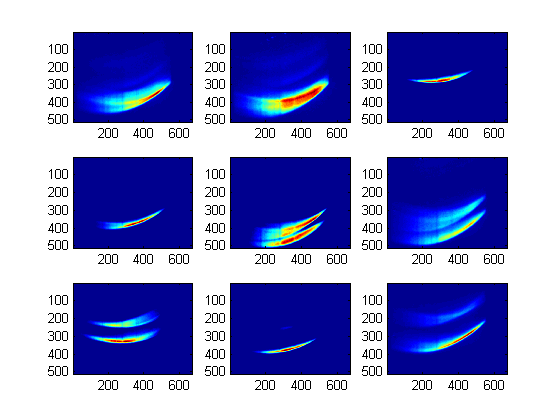}\\
\includegraphics[height=1in]{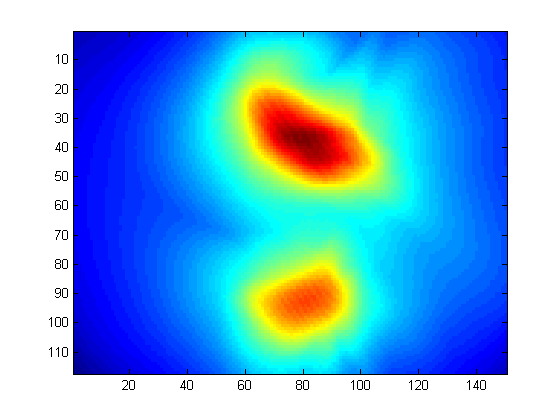}
\includegraphics[height=1in]{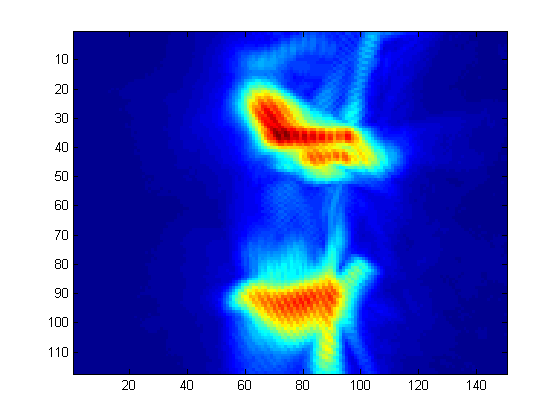}
\includegraphics[height=1in]{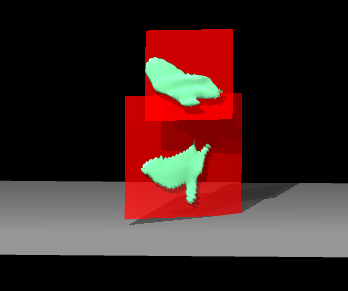}
\includegraphics[height=1in]{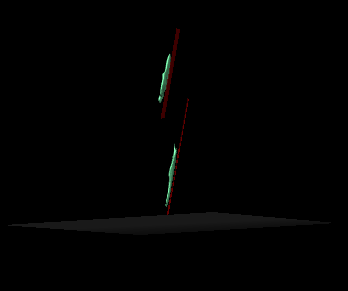}

\caption{\label{fig:flow_diagram}Depiction of our reconstruction algorithm for a scene consisting of two birds in different planes. From top left to bottom right: Photographs of the input models. 9 out of 33 streak images used for reconstruction. The raw (unfiltered) backprojection. The filtered backprojection, after taking a second derivative. 3D renderings in Chimera.}
\end{figure}

We recorded series of streak images for several simple 3D scenes, comprised of white Lambertian objects. We used 30-60 laser positions, spread over a 20 x 40 cm wall. 
The reconstructed surfaces are displayed in Figures \ref{fig:recon_shapes}, \ref{fig:recon_gandalf}, \ref{fig:flow_diagram}. To produce the 3D pictures from the volumetric data, we use the Chimera visualization software \cite{chimerasoftware} for thresholding and rendering.

\begin{figure}
\centering
\includegraphics[height=1in]{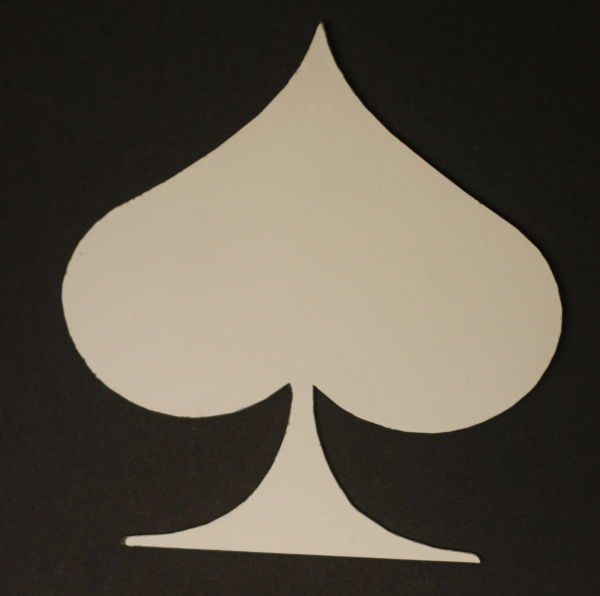}
\includegraphics[height=1in]{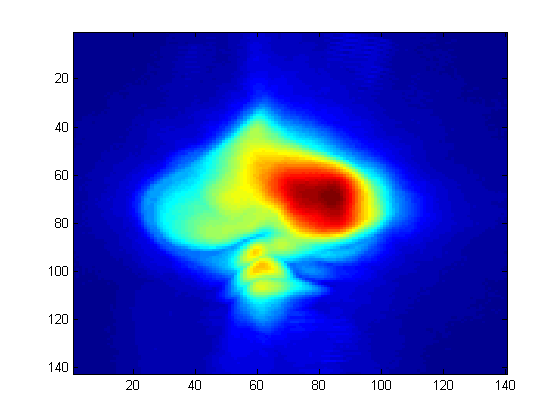}
\includegraphics[height=1in]{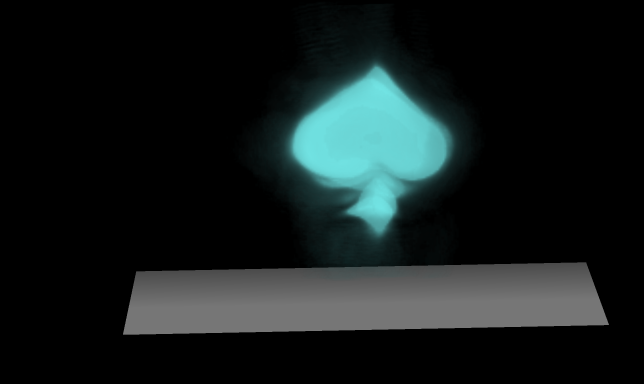}
\caption{\label{fig:recon_spade} Reconstruction of a planar object in an unknown plane in 3D. (Left) The object. (Middle Left) 2D Projection of the filtered heatmap. (Middle Right) A 3D visualization of the filtered heatmap. (Right) Reconstruction using sparsity based methods. The gray ground plane has been added to aid visualization. }
\end{figure}

\subsection{Performance Evaluation}

We conducted several experiments to test the performance of our experimental setup. This includes verifying the spatial and temporal resolution of the camera and the resolution obtained in a simple hidden scene. The resolution in the hidden scene depends greatly on the position in the scene and overall scene complexity. Experiments indicate that for a simple patch a precision of about 500~$\mu$m perpendicular to the wall and 1~cm parallel to the wall is achievable.


\begin{figure}[ht]
\centering
\includegraphics[width=3in]{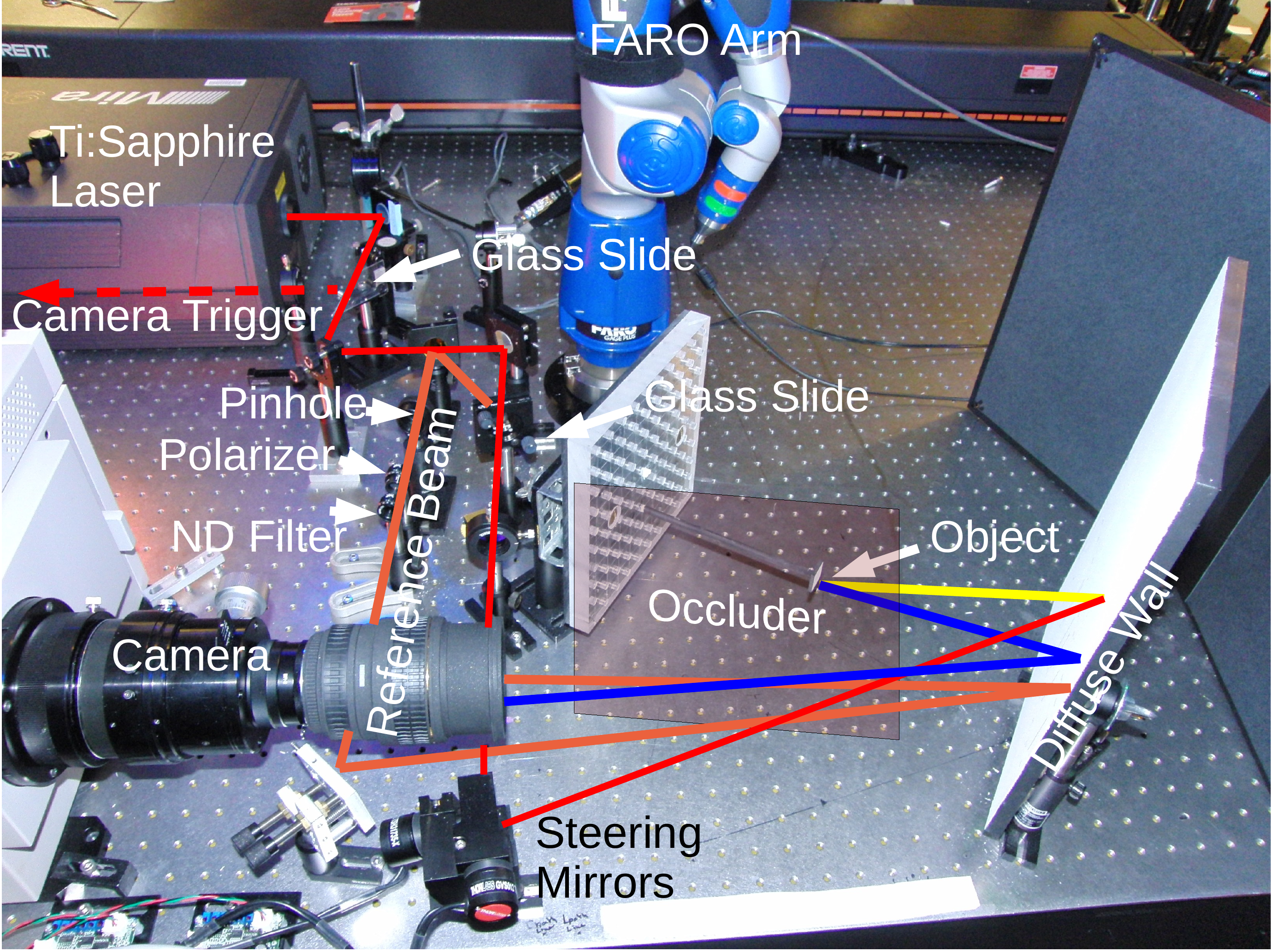}
\caption{\label{fig:setupphoto}The laser beam (red) is split to provide a syncronization signal for the camera (dotted red) and an attenuated reference pulse (orange) to compensate for synchronization drifts and laser intensity fluctiations. The main laser beam is directed to a wall with a steering mirror and the returned third bounce is captured by the streak camera. An Occluder inserted at the indicated position does not significantly change the collected image.}
\end{figure}



%
%
\section{Future Directions}


We have shown that the goal of recovering hidden shapes is only as challenging as the current hardware. The computational approaches show great promise. But, on the hardware front, emerging integrated solid state lasers, new sensors and non-linear optics will provide practical and portable imaging devices. Our formulation is also valid for shorter wavelengths (e.g., x-rays) or for ultrasound and sonar frequencies in large scenes where diffraction can be neglected. Beyond geometry, one maybe able to recover full light transport and bidirectional reflectance distribution function (BRDF) from a single viewpoint to eliminate encircling instrumentation.  Our current method assumes friendly reflectances, i.e., a non-zero diffuse component towards the wall. Non-lambertian reflectance and partial occlusions will create non-uniform angular radiance. While our method is robust towards deviations from a Lambertian reflector, reconstructions will benefit from adjusting the model to account for the particular scenario. Generally, non-Lambertian reflectors provide potential challenges by introducing the reflectance distribution as another unknown into the problem and by increasing the dynamic range of the intensities detected by the system. On the other hand, they may provide the benefit of higher reflected intensities and better defined reflection angles that could be exploited by an adequate reconstruction method. The visible wall need not be planar and one can update the $r_l$ and $r_c$ distances from a known model of the visible parts. Supporting refraction involves multiple or continuous change in the path vector. Atcheson et. al. have shown a refraction tomography approach \cite{atcheson2008time} which could be extended.

Initial applications may be in controlled settings like endoscopy, scientific imaging and industrial vision. This will require addressing more complex transport for volumetric scattering (e.g. for tissue) or refracting elements (e.g. for fluids). A very promising theoretical direction is in inference and inversion techniques that exploit scene priors, sparsity, rank, meaningful transforms and achieve bounded approximations. Adaptive sampling can decide the next-best laser direction based on current estimate of the carved hull. Future analysis will include coded sampling using compressive techniques and noise models for SNR and effective bandwidth. 

%

We used the COSAMP
matching pursuit algorithm which allows us to explore the sparsity of the solution. We tested both the backprojection and linear equation based methods on both artificially generated and real data. It turned out that for artificial data the COSAMP based reconstruction algorithm was generally superior the backprojection algorithm. The backprojection algorithm can recover objects front-to parallel to the wall quite well, but fails for highly sloped surfaces. However, for real data the linear equation based methods were very sensitive to calibration errors, i.e., errors in the matrix $A$ above. One can obtain results for very good datasets, after changing $A$ slightly to account for intricacies of our imaging system like vignetting and gain correction. On the other hand the backprojection algorithm turned out to be quite robust to calibration errors, though they can deteriorate the obtained resolution. Since we typically have to deal with some calibration error (see section \ref{sec:experiments}), we used the backprojection algorithm to obtain the real data reconstruction results of this paper.

Designing a perfectly calibrated system is an interesting research challenge. Research directions here include the development of a model for lens distortions in the streak camera system and a scheme for their compensation, as well as a method of actively and autonomously recalibrating the system to account for day to day variations and drift in laser operation and laser-camera synchronization wile capturing data.

\section{Conclusion}


The ability to infer shapes of objects beyond the line of sight is an ambitious goal but it may transform recoding of visual information and will require a new set of algorithms for scene understanding, rendering and visualization. 
We have presented a new shape-from-x approach that goes beyond the abilities of today's model acquisition and scanning methods. Light transport with a time component presents a unique challenge due to the lack of correspondence but also provides a new opportunity. The emphasis in this paper is to present a forward model and novel inversion process. 
The nonlinear component due to jitter and system point spread function makes the ultra-fast imaging equipment difficult to use. So the physical results can only be treated as a proof-of-concept.

One may wonder when ultrafast lasers and cameras will be broadly available to researchers in computer graphics and computational photography. Many lasers have transformed from their unsafe, bulky form factors to use in portable consumer devices. 
To further the research in this field by the community, we will make image datasets and matlab code freely available online.

The utility of ultrafast imagers like the streak camera has been limited to analysis of bio-chemical processes making them difficult to use for free space light transport. We hope our work will spur more applications of these imagers in computational photography and in turn the graphics research will influence the design and cost of future streak cameras. 



\end{document}